\documentclass[fleqn,usenatbib,useAMS]{mnras}
 \usepackage{graphicx}
 \usepackage{amsmath}
 \usepackage{amssymb}
 \usepackage[T1]{fontenc}
 \usepackage{ae,aecompl}
 \usepackage{txfonts}

 \title[The NEA pair 2017~SN$_{16}$--2018~RY$_{7}$]
       {Dancing with Venus in the shadow of the Earth: a pair of genetically 
        related near-Earth asteroids trapped in a mean-motion resonance}

 \author[C. de la Fuente Marcos and R. de la Fuente Marcos]
        {C.~de~la~Fuente~Marcos$^{1}$\thanks{E-mail: nbplanet@ucm.es}
         and
         R. de la Fuente Marcos$^2$ \\
         $^1$ Universidad Complutense de Madrid,
              Ciudad Universitaria, E-28040 Madrid, Spain \\
         $^2$AEGORA Research Group,
             Facultad de Ciencias Matem\'aticas,
             Universidad Complutense de Madrid,
             Ciudad Universitaria, E-28040 Madrid, Spain}
 \date{Accepted 2018 November 12.
       Received 2018 November 6; 
       in original form 2018 October 1}
 \pubyear{2019}
 
\begin{document}
  \label{firstpage}
  \pagerange{\pageref{firstpage}--\pageref{lastpage}}
  \maketitle

  \begin{abstract}
     YORP-induced fission events may form dynamically coherent pairs or even
     families of asteroids. The outcome of this process is well documented 
     among members of the main asteroid belt, but not in the case of the 
     near-Earth asteroid (NEA) population because their paths randomize very 
     efficiently in a short time-scale. Mean-motion resonances (MMRs) may 
     stabilize the orbits of small bodies by making them avoid close 
     encounters with planets. In theory, YORP-induced fission of asteroids 
     trapped in MMRs can preserve evidence of this process even in near-Earth 
     space. Here, we show that two NEAs, 2017~SN$_{16}$ and 2018~RY$_{7}$, 
     are currently following an orbital evolution in which their relative 
     mean longitude does not exhibit any secular increase due to the 
     stabilizing action of the 3:5 MMR with Venus. The mechanism that makes 
     this configuration possible may be at work both in the Solar system and 
     elsewhere. Our analysis suggests that the pair 2017~SN$_{16}$--2018~RY$_{7}$ 
     may have had its origin in one out of two mechanisms: YORP-induced 
     splitting or binary dissociation. 
  \end{abstract}


  \section{Introduction}
     The Yarkovsky--O'Keefe--Radzievskii--Paddack (YORP) effect (see e.g. \citealt{2006AREPS..34..157B}) can induce spin-up of asteroids and 
     mass shedding. Dynamically coherent pairs or groups of asteroids probably produced by this process have been found among members of the 
     main asteroid belt (see e.g. \citealt{2008AJ....136..280V,2018Icar..304..110P}). Near-Earth asteroids (NEAs) should also fission via 
     the YORP mechanism (see e.g. \citealt{2011Icar..214..161J}), but any dynamically coherent pairs resulting from this process are 
     difficult to identify because the orbits randomize very quickly in near-Earth space (see e.g. \citealt{2012Icar..220.1050S,2014Icar..238..156S}). 

     Mean-motion resonances (MMRs; see e.g. \citealt{2006Icar..184...29G,2019Icar..317..121G}) can make orbits long-term stable by protecting 
     small bodies against close encounters with planets, in our case the Earth--Moon system (see e.g. \citealt{1989Icar...78..212M}). In 
     theory, YORP-induced fission of asteroids trapped in MMRs may preserve the evidence of this process even in near-Earth space. Resonant 
     confinement has been previously discussed within the context of cometary dust dynamics (see e.g. \citealt{1999MNRAS.304L..53A}) and 
     ring dynamics (see e.g. \citealt{2002Natur.417...45N}). Here, we show that two NEAs, 2017~SN$_{16}$ and 2018~RY$_{7}$, are currently 
     trapped in the 3:5 MMR with Venus and following an unusual mutual orbital evolution, which may be consistent with an origin in a 
     YORP-induced fission event. This Letter is organized as follows. Section~2 presents the available data on this pair of NEAs. Their 
     orbital evolution is studied in Section~3 and their origin discussed in Section~4. In Section 5, we compare with predictions from a new 
     four-dimensional orbit model of the NEA population. Our results are discussed in Section 6. Section 7 summarizes our conclusions.
%
%
     \begin{table*}
      \centering
      \fontsize{8}{11pt}\selectfont
      \tabcolsep 0.35truecm
      \caption{Heliocentric Keplerian orbital elements and associated 1$\sigma$ uncertainties of 2010~AF$_{3}$, 2017~SN$_{16}$, and 
               2018~RY$_{7}$. The orbit determination of 2017~SN$_{16}$ was computed on 2018 November 3 and it is based on 97 astrometric 
               observations for a data-arc span of 391 d; the one of 2018~RY$_{7}$ was computed on 2018 November 3 and it is based on 74 
               observations for a data-arc span of 36 d; the one of 2010~AF$_{3}$ was computed on 2017 April 6 and it is based on 26 
               observations for a data-arc span of 7 d. Orbit determinations are referred to epoch JD 2458600.5 (2019-Apr-27.0) TDB (J2000.0 
               ecliptic and equinox). Source: JPL's SBDB.
              }
      \begin{tabular}{lcccc}
       \hline
        Orbital parameter                                 &   & 2010~AF$_{3}$       & 2017~SN$_{16}$              & 2018~RY$_{7}$         \\
       \hline
        Semimajor axis, $a$ (au)                          & = &   1.0166$\pm$0.0002 &   1.01613704$\pm$0.00000004 &   1.01616$\pm$0.00003 \\
        Eccentricity, $e$                                 & = &   0.1236$\pm$0.0004 &   0.1455151$\pm$0.0000005   &   0.14699$\pm$0.00005 \\
        Inclination, $i$ (\degr)                          & = &  11.82$\pm$0.04     &  13.38253$\pm$0.00003       &  13.348$\pm$0.007     \\
        Longitude of the ascending node, $\Omega$ (\degr) & = & 285.644$\pm$0.005   &   2.732361$\pm$0.000010     &   2.81708$\pm$0.00009 \\
        Argument of perihelion, $\omega$ (\degr)          & = & 289.959$\pm$0.013   & 137.97946$\pm$0.00011       & 136.879$\pm$0.004     \\
        Mean anomaly, $M$ (\degr)                         & = & 293.2$\pm$1.1       &  77.9182$\pm$0.0002         &  80.4804$\pm$0.0006   \\
        Perihelion, $q$ (au)                              & = &   0.8909$\pm$0.0002 &   0.8682738$\pm$0.0000005   &   0.86680$\pm$0.00002 \\
        Aphelion, $Q$ (au)                                & = &   1.1422$\pm$0.0003 &   1.16400030$\pm$0.00000005 &   1.16553$\pm$0.00006 \\
        Absolute magnitude, $H$ (mag)                     & = &  26.1               &  23.3                       &  24.4                 \\
       \hline
      \end{tabular}
      \label{elements}
     \end{table*}
%
%

  \section{The NEA pair 2017~SN$_\mathbf{16}$--2018~RY$_\mathbf{7}$: data}
     The orbit determinations used in this work have been obtained from Jet Propulsion Laboratory's (JPL) Small-Body Database 
     (SBDB).\footnote{\href{https://ssd.jpl.nasa.gov/sbdb.cgi}{https://ssd.jpl.nasa.gov/sbdb.cgi}} Minor body 2017~SN$_{16}$ was discovered 
     by A. R. Gibbs working for the Mount Lemmon Survey in Arizona (1.5-m reflector + 10K CCD) on 2017 September 24 \citep{2017MPEC....S..186S}. 
     Additional data have been obtained during the last favourable observation window \citep{2018MPEC....R...38G}. It is a relatively small 
     object with an absolute magnitude, $H=23.3$~mag (assumed $G$ = 0.15), which suggests a diameter close to 80~m, but with a possible 
     range of values of 38--170~m for an assumed albedo in the range 0.60--0.03. This Apollo asteroid has a semimajor axis $a$ = 1.0161~au, 
     and moves in a low-eccentricity, $e$ = 0.1455, and moderate-inclination, $i=$13\fdg38, orbit that keeps it confined to the 
     neighbourhood of the Earth--Moon system (see Table~\ref{elements}); its Minimum Orbit Intersection Distance (MOID) with our planet 
     is 0.093~au. These orbital properties make it relatively easy to access from the Earth and it is part of the Near-Earth Object Human 
     Space Flight Accessible Targets Study (NHATS)\footnote{\href{http://neo.jpl.nasa.gov/nhats/} {http://neo.jpl.nasa.gov/nhats/}} list 
     \citep{2012DPS....4411101A}. Asteroid 2018~RY$_{7}$ was first observed on 2018 September 14 by B. M. Africano also working for the 
     Mount Lemmon Survey \citep{2018MPEC....S...12R}. Its orbit determination still requires improvement, but the values of its orbital 
     elements are markedly similar to those of 2017~SN$_{16}$ (see Table~\ref{elements}); with $H=24.4$~mag it could be $\sim$45~m wide, its 
     MOID with the Earth is 0.094~au, and it is also listed by NHATS. 

     Neglecting binaries and higher multiplicity systems, the pair 2017~SN$_{16}$--2018~RY$_{7}$ shows the highest degree of orbital 
     coherence ever observed among NEAs. Although they are not binary companions, the asteroids happen to be rather close to each other, far
     closer than might be attributted to chance. Such an arrangement has never before been observed among low-mass minor bodies in 
     near-Earth space. Being small NEAs, they may be pieces of larger asteroids and there are a number of processes that can make this 
     possible. In addition to the YORP mechanism pointed out above, subcatastrophic collisions in which a small body hits a larger object 
     can produce fragments (see e.g. \citealt{2007Icar..186..498D}), but they can also be released as a result of tidal disruption events 
     during very close encounters with planets (see e.g. \citealt{2014Icar..238..156S}). On the other hand, the present-day values of their 
     semimajor axes are close to 1.0168037~au, the location of the 3:5 MMR with Venus, and they are both strong candidates to being locked 
     in this planetary resonance. It is however possible that being locked in resonance with Venus plays a major role in preserving their 
     high degree of orbital coherence. In order to validate these theoretical expectations, a representative set of control orbits must be 
     integrated forward and backwards in time to confirm that the dynamical evolution of this pair of NEAs over a reasonable amount of time 
     is consistent with not being close by chance and that the MMR with Venus is actually responsible for what is being observed. The 
     critical angles relevant to such a numerical exploration are the relative mean longitude, $\lambda_{\rm r}$, or difference between the 
     mean longitudes of the NEAs (to study the mutual evolution of the pair), and the resonant angle between one NEA and Venus, 
     $\sigma_{\rm V}=5\lambda-3\lambda_{\rm V}-2 (\Omega+\omega)$ ---to study the 3:5 MMR with Venus. In celestial mechanics, the mean 
     longitude of an object ---planet or minor body--- is given by $\lambda=M+\Omega+\omega$, where $M$ is the mean anomaly, $\Omega$ is the 
     longitude of the ascending node, and $\omega$ is the argument of perihelion (see e.g. \citealt{1999ssd..book.....M}). Resonance happens 
     when the value of $\sigma_{\rm V}$ oscillates or librates over time. The 3:5 MMR with Venus is not the strongest or traditionally most 
     populated \citep{2006Icar..184...29G}. 

  \section{The NEA pair 2017~SN$_\mathbf{16}$--2018~RY$_\mathbf{7}$: evolution}
\begin{figure}
        \centering
        \includegraphics[width=\linewidth]{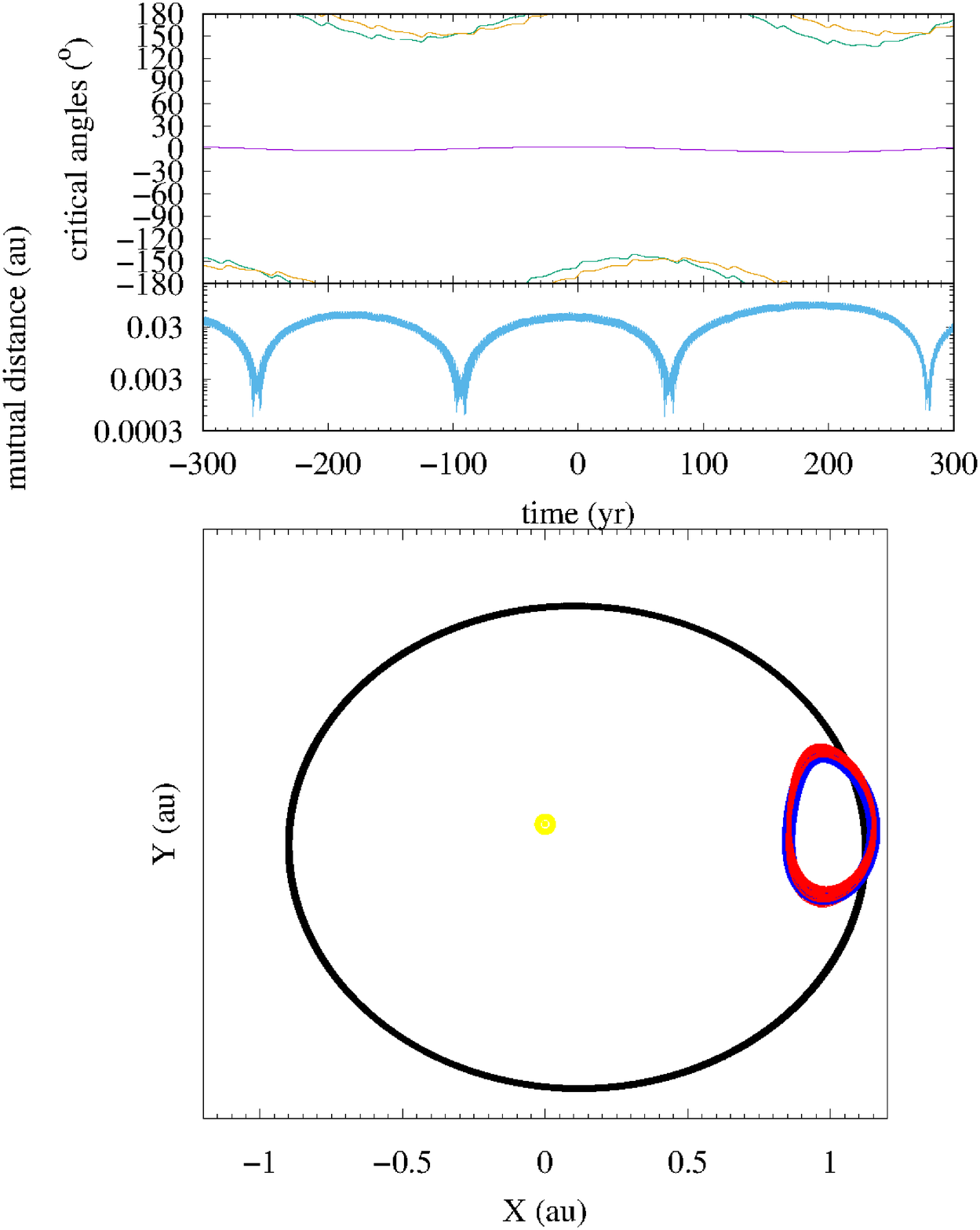}
        \caption{The top panel shows the evolution of the values of the critical angles over the time span ($-$300, 300)~yr according to the 
                 nominal orbit determinations in Table \ref{elements} ---in purple, the relative mean longitude of the pair 
                 2017~SN$_{16}$--2018~RY$_{7}$, in orange, the resonant angle associated with the 3:5 MMR of 2017~SN$_{16}$ with Venus, and
                 in teal, the one of 2018~RY$_{7}$. The mutual distance is displayed in the middle panel. The bottom panel shows the orbital 
                 arrangement projected on to the ecliptic plane in a frame of reference centred at the Sun that rotates with 2017~SN$_{16}$ 
                 ---2017~SN$_{16}$ in blue, 2018~RY$_{7}$ in red. The orbit of 2017~SN$_{16}$ is indicated in black. All the control orbits 
                 investigated in this work evolve in a similar fashion within this time interval. The zero-point in time corresponds to 
                 epoch JD~2458600.5~TDB, 27-April-2019. 
                }
        \label{pairevolution}
     \end{figure}

\begin{figure}
        \centering
         \includegraphics[width=\linewidth]{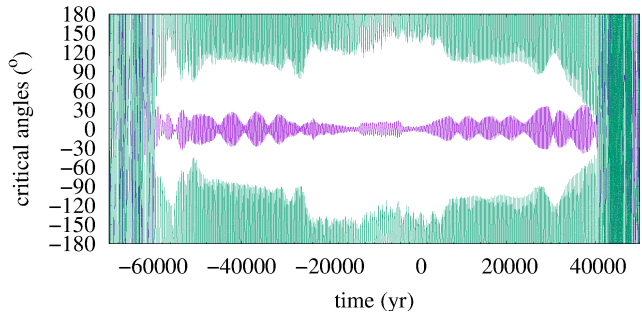}
         \caption{Same as Fig.~\ref{pairevolution}, top panel, but showing the evolution of the critical angles over a longer period of 
                  time. 
                 }
         \label{resonances}
      \end{figure}

     In order to explore the details of the orbital evolution of the pair 2017~SN$_{16}$--2018~RY$_{7}$, we use a direct $N$-body code that 
     implements a fourth-order version of the Hermite integration scheme \citep{1991ApJ...369..200M,2003gnbs.book.....A}. The standard 
     version of this software is publicly available from the IoA web site.\footnote{\href{http://www.ast.cam.ac.uk/~sverre/web/pages/nbody.htm}
     {http://www.ast.cam.ac.uk/$\sim$sverre/web/pages/nbody.htm}} Our calculations use the latest orbit determinations and include 
     perturbations by the eight major planets, the Moon, the barycentre of the Pluto--Charon system, and the three largest asteroids. 
     Further details of the code and of our overall approach and physical model can be found in \citet{2012MNRAS.427..728D}. Data to 
     generate initial conditions as well as other input data have been obtained from JPL's SBDB. Figure~\ref{pairevolution} shows the 
     short-term orbital evolution of the pair (nominal orbits in Table~\ref{elements}) and confirms that 2017~SN$_{16}$ and 2018~RY$_{7}$ 
     are engaged in an unusual dance that keeps them not far from each other for an extended period of time. As suspected, the 3:5 MMR of 
     the pair with Venus keeps them together (see Fig.~\ref{resonances}). When the pair leaves the planetary resonance their dancing 
     engagement ends abruptly. It can however be argued that the orbit of 2018~RY$_{7}$ is too uncertain to confirm this analysis. 
%
%
%
%
%
%
%
%

\begin{figure}
        \centering
        \includegraphics[width=\linewidth]{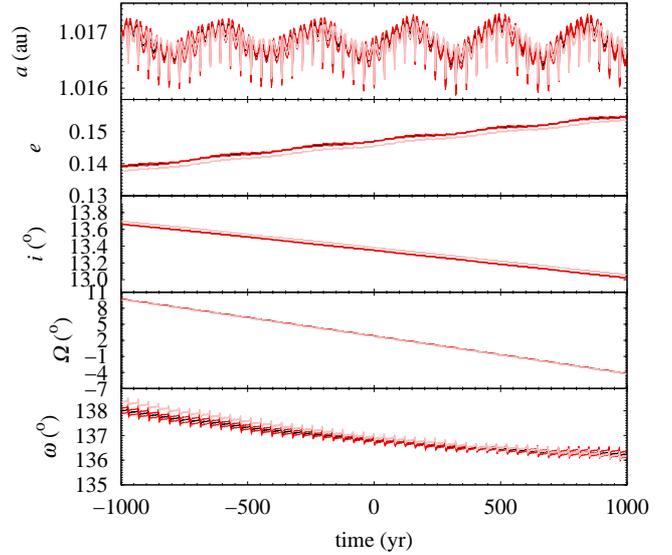}
        \caption{Time evolution of the dispersions of the values of the orbital elements of 2017~SN$_{16}$ (grey/pink) and 2018~RY$_{7}$ 
                 (black/red): semimajor axis (top panel), eccentricity (second to top panel), inclination (middle panel), longitude of the 
                 ascending node (second to bottom panel), and argument of perihelion (bottom panel). Average values are displayed as thick 
                 grey/black curves and their ranges (1$\sigma$ uncertainties) as thin pink/red curves.  
                }
        \label{disper}
     \end{figure}

\begin{figure}
        \centering
         \includegraphics[width=\linewidth]{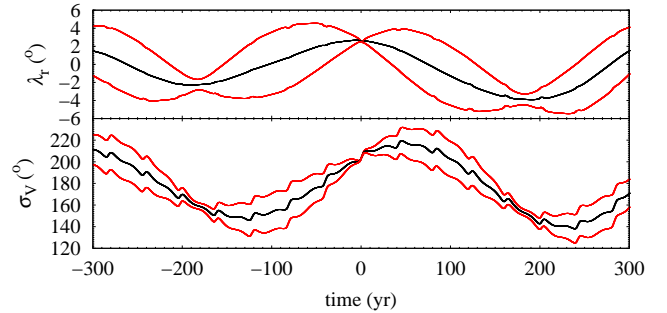}
         \caption{Evolution of the dispersions (in red, mean in black) of the value of the relative mean longitude (top panel) of the 
                  pair 2017~SN$_{16}$--2018~RY$_{7}$ and that of the resonant angle (bottom panel) of 2018~RY$_{7}$ with Venus.  
                 }
         \label{lambdar}
      \end{figure}

     In order to investigate if the uncertainties have an impact on our results, we have applied the covariance matrix methodology described 
     in \citet{2015MNRAS.453.1288D}; the covariance matrices necessary to generate initial conditions have been obtained from JPL's SBDB. 
     The results of the evolution of 500 control orbits generated using this approach are presented in Fig.~\ref{disper}; in general, the 
     dispersions (in pink and red) are too small to play any role. Our analysis also suggests that the dynamical age of this pair is younger 
     than about 60\,000~yr (see Fig.~\ref{resonances}), although the most likely value is around 14\,600 yr (not shown in the figures). If 
     we focus on the critical angles, Fig.~\ref{lambdar} shows the results of 1000 control orbits and the dispersions are consistently 
     small. Therefore, we can confirm that our numerical results are robust and the orbital evolution of this pair is well characterized 
     within the time window explored here. 
%
%
%
%
%
%
%
%

  \section{The NEA pair 2017~SN$_\mathbf{16}$--2018~RY$_\mathbf{7}$: origin}
     Regarding the origin of the pair 2017~SN$_{16}$--2018~RY$_{7}$, four scenarios may be considered: accidental proximity induced by 
     differential precession in $\Omega$ and $\omega$, tidal disruption after a close flyby with our planet, binary dissociation, and 
     YORP-induced rotational disruption. Although the first mechanism appears to be behind the orbital evolution of 15810 Arawn 
     (1994~JR$_1$) ---see \citet{2012MNRAS.427L..85D, 2016MNRAS.462.3344D} and \citet{2016ApJ...828L..15P}--- it can be easily discarded in 
     the present case because Fig.~\ref{resonances} shows that the orbital engagement between 2017~SN$_{16}$ and 2018~RY$_{7}$ is not of a 
     recurrent nature, but a configuration that remains relatively stable for an extended period of time (see the value of the relevant 
     critical angle, in purple); Arawn could be a quasi-satellite of Pluto, but this is not a plausible dynamical status for the pair of 
     NEAs under analysis here (more on this in Section 6). Tidal disruption after a planetary close encounter must be discarded as well 
     because flybys at rather short planetary distances, in the range 2--5 planetary radii (see e.g. \citealt{1992Icar...95...86S}), are 
     required and this is not observed during the stable phase of the simulations. As for the third scenario, binary dissociation requires 
     the presence of a pre-existing binary system, but 2017~SN$_{16}$ has $H=23.2$~mag and one may wonder if asteroids that small may host 
     long-term binary companions. At the time of this writing, the faintest known asteroid with a companion is the NEA 2015~TD$_{144}$ with 
     $H=22.6$~mag;\footnote{\href{https://echo.jpl.nasa.gov/~lance/binary.neas.html}{https://echo.jpl.nasa.gov/~lance/binary.neas.html}} 
     therefore, it is in principle possible that 2018~RY$_{7}$ could be a former binary companion of 2017~SN$_{16}$ that became unbound at 
     some point in the past as a result of an external action (e.g. small impact). The asteroidal YORP effect ---which is the result of 
     anisotropic reemission of sunlight from the surfaces of the affected minor bodies--- can slowly increase their rotational speed, 
     leading them to reach their fission limit and eventual disruption (see e.g. \citealt{2012Icar..220..514W,2016Icar..277..381J}). This 
     mechanism is behind the fourth scenario and it can produce binaries as well as unbound asteroid pairs. With the available data, it is 
     virtually impossible to decide: 2018~RY$_{7}$ may well be a YORPlet as described by \citet{2017DPS....4930205C} that came from the 
     putative disrupted progenitor of 2017~SN$_{16}$, but it may also be a former binary companion of 2017~SN$_{16}$ formed by YORP-induced 
     fission, or some other mechanism, that became unbound at some point. 

  \section{NEO orbit model predictions}
\begin{figure}
        \centering
         \includegraphics[width=\linewidth]{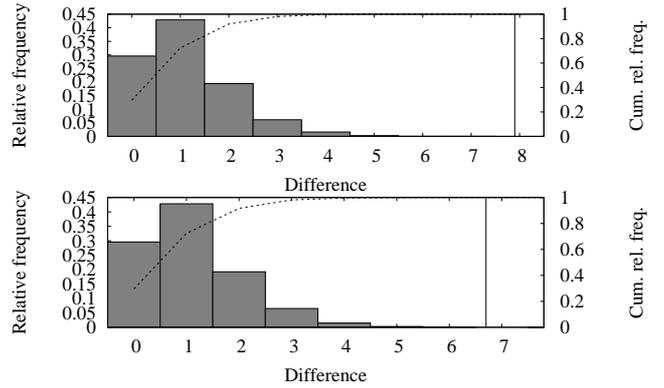}
         \caption{Results of the randomization test described in the text, 2017~SN$_{16}$ (top panel) and 2018~RY$_{7}$ (bottom panel). The
                  values of the test statistics are represented by vertical lines. 
                 }
         \label{randomization}
      \end{figure}

     If the NEA pair 2017~SN$_{16}$--2018~RY$_{7}$ (and perhaps other NEAs in similar orbits) is the result of a recent fragmentation or 
     binary dissociation event, such orbits must be largely absent from synthetic, debiased data from a near-Earth object (NEO) population 
     model that includes both asteroid and comets. By comparing observational and synthetic data, we may be able to understand better the 
     circumstances surrounding the formation of this unusual NEA pair. The NEO orbit model developed by the Near-Earth Object Population 
     Observation Program (NEOPOP) and described by \citet{2018Icar..312..181G} is the state-of-the-art tool fit for the purpose; this new 
     four-dimensional model provides debiased steady-state distributions of $a$, $e$, $i$, and $H$ for $H<25$~mag. The software that 
     implements this model is publicly available\footnote{\href{http://neo.ssa.esa.int/neo-population}{http://neo.ssa.esa.int/neo-population}} 
     and it has been successfully validated \citep{2016Natur.530..303G,2017A&A...598A..52G,2018Icar..311..271G}. We have used the list of 
     NEOs with $H<25$~mag currently catalogued (as of 2018 November 4, 14\,390 objects) to estimate how likely is that the pair 
     2017~SN$_{16}$--2018~RY$_{7}$ could be explained by the NEO orbit model. In order to do this, we have applied a randomization test 
     \citep{1970smrw.book.....F}. As test statistics, we use the differences between the observed number of NEOs in orbits close to those of 
     the pair 2017~SN$_{16}$--2018~RY$_{7}$ and the number predicted by the NEOPOP software. Following \citet{2018MNRAS.473.3434D}, we 
     consider two $D$-criteria, $D_{\rm LS}$ and $D_{\rm R} < 0.05$, to count two NEOs as dynamically similar ---$D_{\rm LS}$ as in 
     equation 1 of \citet{1994ASPC...63...62L} and the $D_{\rm R}$ from \citet{1999MNRAS.304..743V}. We find nine NEOs that follow 
     2017~SN$_{16}$-like orbits and eight that follow 2018~RY$_{7}$-like ones; in contrast, NEOPOP predicts 1.1$\pm$0.9 and 1.3$\pm$1.1, 
     respectively, for a synthetic sample of the same size. With these results, the respective differences (our test statistics) are 
     7.9$\pm$0.9 and 6.7$\pm$1.0. If we extract two random samples, we can compute the number of synthetic NEOs in orbits similar to each 
     member of the pair for both samples and calculate the differences. In order to obtain statistically significant results, we have 
     repeated this experiment 10\,000 times and our results are summarized in Fig.~\ref{randomization} where our test statistics are plotted 
     as vertical black lines. The probability of obtaining a difference $>7.0$ for 2017~SN$_{16}$-like orbits is 0.0001 and that of getting 
     a value $>7.9$ is 0.0001; for 2018~RY$_{7}$, the probability of obtaining a value $>5.7$ is 0.0011 and that of getting one $>6.7$ is 
     0.0001. Our analysis using NEOPOP strongly suggests that this pair of NEOs may have not followed the conventional dynamical pathways 
     that populate near-Earth orbital parameter space. The pair must have been produced within near-Earth space. 
%
%
%
%

  \section{Discussion}
\begin{figure}
        \centering
         \includegraphics[width=\linewidth]{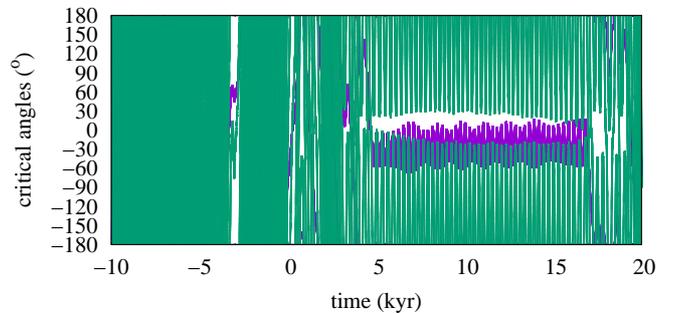}
         \caption{Similar to Fig.~\ref{resonances} but for 2017~SN$_{16}$ and 2010~AF$_{3}$, another NEA that is in near 3:5 MMR with Venus. 
                  In teal, the resonant angle associated with the 3:5 MMR of 2010~AF$_{3}$ with Venus. An episode of dynamical coherence (of 
                  2010~AF$_{3}$ with respect to 2017~SN$_{16}$) is visible. 
                 }
         \label{resonances2}
      \end{figure}

     The study of the pair of NEAs considered here opens a window into the present-day dynamical processes that are shaping the NEO 
     population. Even if most NEOs may have been originally delivered from the main asteroid belt, their arrival parameters do not remain 
     frozen in time and a tangled web of mean-motion and secular resonances (see e.g. \citealt{2016MNRAS.456.2946D}) together with the 
     Yarkovsky and YORP effects, and perhaps others, might continuously modify the distributions of $a$, $e$, $i$, and $H$. This is at least 
     what can be understood from the study of the NEA pair 2017~SN$_{16}$--2018~RY$_{7}$. The unusual dynamical behaviour observed may not 
     be exclusive of this pair; in Fig.~\ref{resonances2} we observe another episode of dynamical coherence, in this case of 2010~AF$_{3}$ 
     ---another NEA close to the 3:5 MMR with Venus--- with respect to 2017~SN$_{16}$, in the future. 
%
%
%
%
     
     Although the value of the relative mean longitude of the pair 2017~SN$_{16}$--2018~RY$_{7}$ oscillates over time around 0\degr, these 
     objects are not engaged in a mutual quasi-satellite dynamical state. In the Solar system, quasi-satellites are minor bodies that appear 
     to travel retrograde or backwards around a host when observed in a frame of reference that rotates with the host (see e.g. 
     \citealt{2006MNRAS.369...15M,2016MNRAS.462.3344D}). The value of the relative mean longitude of the quasi-satellite with respect to the 
     host librates, but the orbital evolution of the quasi-satellite is controlled by the combined action of the Sun and the host. The 
     probable values of the masses of the NEAs studied here are too small to play any role. Because of this, the case of this pair of NEAs 
     is very different from that of Saturn and the moons Janus and Epimetheus (see e.g. \citealt{1999ssd..book.....M}). 
     Figure~\ref{pairevolution} can be explained as the result of two asteroids being concurrently trapped in the 3:5 MMR with Venus. Here, 
     the average angular speed of the asteroids is nearly the same and equal to that which corresponds to the resonance location. Due to 
     this, the difference in the mean longitudes of the asteroids shown in Figs~\ref{pairevolution} (top panel), \ref{resonances}, and 
     \ref{lambdar} does not exhibit any secular increase and thus resembles what is seen for objects trapped in the quasi-satellite state. 
     What we have shown is that two NEAs are currently engaged in a faux-binary configuration, thanks to the stabilizing action of the 3:5 
     MMR with Venus. The essence of this mechanism is summarized in Fig.~\ref{resonances}, where the libration of $\lambda_{\rm r}$ is 
     sustained by that of $\sigma_{\rm V}$. The mechanism that makes this configuration possible may be at work elsewhere as long as all the 
     ingredients are present: small bodies trapped in a stable planetary MMR. 

  \section{Conclusions}
     In this Letter, we have presented the first example of a new type of orbital configuration, a pair of asteroids kept close to each 
     other for an extended period of time by a non-co-orbital MMR. This study has been carried out using the latest data, direct $N$-body 
     calculations, a state-of-the-art NEO orbit model, and statistical analyses. Our conclusions can be summarized as follows:
     \begin{enumerate}
        \item We have identified a pair of NEAs, 2017~SN$_{16}$--2018~RY$_{7}$, trapped in the 3:5 MMR with Venus that
              seem to orbit around a common point when viewed in a frame of reference co-rotating with the pair. Their evolution resembles
              that of a quasi-satellite, but they are not engaged in true quasi-satellite resonant behaviour as the values of their masses 
              are negligible.  
        \item Extensive calculations show that the pair of NEAs 2017~SN$_{16}$--2018~RY$_{7}$ may have been engaged in its current orbital
              dance for several thousands of years and they will remain in the same dynamical state for a similar amount of time. 
        \item Mechanisms able to create such a peculiar pair include YORP-induced splitting and binary dissociation within MMRs; simple 
              resonance trapping cannot explain the high degree of orbital coherence exhibited by the pair of NEAs
              2017~SN$_{16}$--2018~RY$_{7}$. Given the nature of the past orbital evolution of this pair, its existence is perhaps the first 
              piece of solid evidence in favour of YORP-induced rotational disruption (or binary dissociation) taking place in the immediate 
              neighbourhood of our planet.
        \item The orbital configuration studied here may also be found among other small bodies trapped in MMRs both in near-Earth 
              space and elsewhere.
     \end{enumerate}
     This unusual pair of NEAs will remain favourably positioned for further investigation during the next few years. Spectroscopic studies 
     of 2017~SN$_{16}$--2018~RY$_{7}$ during their future approaches to our planet (2019--2022) should be able to confirm whether they have 
     similar chemical compositions or not, and therefore shed additional light on the mechanism that led to their formation. New data (e.g. 
     astrometry, light curves, and albedos) may also clarify the role of the Yarkovsky and YORP effects within the context of this 
     particularly complex orbital configuration.

  \section*{Acknowledgements}
     We thank the anonymous referee for a particularly constructive critique of the first version of this work that led to a very 
     substantial improvement of its contents and for additional comments, S.~J. Aarseth for providing the code used in this research, and 
     A.~I. G\'omez de Castro for providing access to computing facilities. This work was partially supported by the Spanish `Ministerio de 
     Econom\'{\i}a y Competitividad' (MINECO) under grant ESP2015-68908-R. In preparation of this Letter, we made use of the NASA 
     Astrophysics Data System and the MPC data server.

  \bsp
  \label{lastpage}

\begin{thebibliography}{99}
     \bibitem[\protect\citeauthoryear{Aarseth}{2003}]{2003gnbs.book.....A} Aarseth S.~J., 2003,
             Gravitational N-body simulations.
             Cambridge Univ. Press, Cambridge, p.\ 27
     \bibitem[\protect\citeauthoryear{Abell et al.}{2012}]{2012DPS....4411101A} Abell P. A. et~al., 2012,
             AAS/Div. Planet. Sci. Meeting Abstr., 44, 111.01
     \bibitem[\protect\citeauthoryear{Asher, Bailey \& Emel'Yanenko}{1999}]{1999MNRAS.304L..53A} Asher D.~J., Bailey M.~E., 
             Emel'Yanenko V.~V., 1999, 
             MNRAS, 304, L53
     \bibitem[\protect\citeauthoryear{Bottke et al.}{2006}]{2006AREPS..34..157B} Bottke W.~F., Jr., Vokrouhlick{\'y} D., Rubincam D.~P., 
             Nesvorn{\'y} D., 2006,
             Annu. Rev. Earth Planet. Sci., 34, 157
     \bibitem[\protect\citeauthoryear{Christou et al.}{2017}]{2017DPS....4930205C} Christou A., Borisov G., Jacobson S.~A., Colas F., 
             dell'Oro A., Cellino A., Bagnulo S., 2017, 
             AAS/Div. Planet. Sci. Meeting Abstr., 49, 302.05
     \bibitem[\protect\citeauthoryear{de la Fuente Marcos \& de la Fuente Marcos}{2012a}]{2012MNRAS.427..728D} de la Fuente Marcos C.,
             de la Fuente Marcos R., 2012a,
             MNRAS, 427, 728
     \bibitem[\protect\citeauthoryear{de la Fuente Marcos \& de la Fuente Marcos}{2012b}]{2012MNRAS.427L..85D} de la Fuente Marcos C., 
             de la Fuente Marcos R., 2012b, 
             MNRAS, 427, L85
     \bibitem[\protect\citeauthoryear{de la Fuente Marcos \& de la Fuente Marcos}{2015}]{2015MNRAS.453.1288D} de la Fuente Marcos C.,
             de la Fuente Marcos R., 2015,
             MNRAS, 453, 1288
     \bibitem[\protect\citeauthoryear{de la Fuente Marcos \& de la Fuente Marcos}{2016a}]{2016MNRAS.456.2946D} de la Fuente Marcos C., 
             de la Fuente Marcos R., 2016a, 
             MNRAS, 456, 2946
     \bibitem[\protect\citeauthoryear{de la Fuente Marcos \& de la Fuente Marcos}{2016b}]{2016MNRAS.462.3344D} de la Fuente Marcos C., 
             de la Fuente Marcos R., 2016b, 
             MNRAS, 462, 3344
     \bibitem[\protect\citeauthoryear{de la Fuente Marcos \& de la Fuente Marcos}{2018}]{2018MNRAS.473.3434D} de la Fuente Marcos C., 
             de la Fuente Marcos R., 2018, 
             MNRAS, 473, 3434
     \bibitem[\protect\citeauthoryear{Durda et al.}{2007}]{2007Icar..186..498D} Durda D.~D., Bottke W.~F., Nesvorn{\'y} D., Enke B.~L., 
             Merline W.~J., Asphaug E., Richardson D.~C., 2007, 
             Icarus, 186, 498
     \bibitem[\protect\citeauthoryear{Fisher}{1935}]{1970smrw.book.....F} Fisher R.~A., 1935,
             The design of experiments.
             Oliver and Boyd, Edinburgh
     \bibitem[\protect\citeauthoryear{Gallardo}{2006}]{2006Icar..184...29G} Gallardo T., 2006, 
             Icarus, 184, 29
     \bibitem[\protect\citeauthoryear{Gallardo}{2019}]{2019Icar..317..121G} Gallardo T., 2019, 
             Icarus, 317, 121
     \bibitem[\protect\citeauthoryear{Gilmore et al.}{2018}]{2018MPEC....R...38G} Gilmore A.~C. et al., 2018, 
             MPEC Circ., MPEC 2018-R38
     \bibitem[\protect\citeauthoryear{Granvik \& Brown}{2018}]{2018Icar..311..271G} Granvik M., Brown P., 2018,
             Icarus, 311, 271
     \bibitem[\protect\citeauthoryear{Granvik et al.}{2016}]{2016Natur.530..303G} Granvik M. et al., 2016,
             Nature, 530, 303
     \bibitem[\protect\citeauthoryear{Granvik et al.}{2017}]{2017A&A...598A..52G} Granvik M., Morbidelli A., Vokrouhlick{\'y} D., Bottke W.~F.,
             Nesvorn{\'y} D., Jedicke R., 2017,
             A\&A, 598, A52
     \bibitem[\protect\citeauthoryear{Granvik et al.}{2018}]{2018Icar..312..181G} Granvik M. et al., 2018,
             Icarus, 312, 181
     \bibitem[\protect\citeauthoryear{Jacobson \& Scheeres}{2011}]{2011Icar..214..161J} Jacobson S.~A., Scheeres D.~J., 2011, 
             Icarus, 214, 161
     \bibitem[\protect\citeauthoryear{Jacobson et al.}{2016}]{2016Icar..277..381J} Jacobson S.~A., Marzari F., Rossi A., Scheeres D.~J., 2016,
             Icarus, 277, 381
     \bibitem[\protect\citeauthoryear{Lindblad}{1994}]{1994ASPC...63...62L} Lindblad B.~A., 1994,
             in Kozai Y., Binzel R. P., Hirayama T., eds, ASP Conf. Ser. Vol.\ 63,
             Seventy-five Years of Hirayama Asteroid Families: the Role of Collisions in the Solar System History.
             Astron. Soc. Pac., San Francisco, p.\ 62
     \bibitem[\protect\citeauthoryear{Makino}{1991}]{1991ApJ...369..200M} Makino J., 1991,
             ApJ, 369, 200
     \bibitem[\protect\citeauthoryear{Mikkola et al.}{2006}]{2006MNRAS.369...15M} Mikkola S., Innanen K., Wiegert P., Connors M., Brasser R., 2006, 
             MNRAS, 369, 15
     \bibitem[\protect\citeauthoryear{Milani et al.}{1989}]{1989Icar...78..212M} Milani A., Carpino M., Hahn G., Nobili A.~M., 1989, 
             Icarus, 78, 212
     \bibitem[\protect\citeauthoryear{Murray \& Dermott}{1999}]{1999ssd..book.....M} Murray C.~D., Dermott S.~F., 1999,
             Solar System Dynamics,
             Cambridge Univ. Press, Cambridge
     \bibitem[\protect\citeauthoryear{Namouni \& Porco}{2002}]{2002Natur.417...45N} Namouni F., Porco C., 2002, 
             Nature, 417, 45
     \bibitem[\protect\citeauthoryear{Porter et al.}{2016}]{2016ApJ...828L..15P} Porter S.~B. et al., 2016, 
             ApJ, 828, L15
     \bibitem[\protect\citeauthoryear{Pravec et al.}{2018}]{2018Icar..304..110P} Pravec P. et al., 2018, 
             Icarus, 304, 110
     \bibitem[\protect\citeauthoryear{Ries et al.}{2018}]{2018MPEC....S...12R} Ries J.~G. et al., 2018, 
             MPEC Circ., MPEC 2018-S12
     \bibitem[\protect\citeauthoryear{Schunov{\'a} et al.}{2012}]{2012Icar..220.1050S} Schunov{\'a} E., Granvik M., Jedicke R., Gronchi G., 
             Wainscoat R., Abe S., 2012, 
             Icarus, 220, 1050
     \bibitem[\protect\citeauthoryear{Schunov{\'a} et al.}{2014}]{2014Icar..238..156S} Schunov{\'a} E., Jedicke R., Walsh K.~J., Granvik M., 
             Wainscoat R.~J., Haghighipour N., 2014, 
             Icarus, 238, 156
     \bibitem[\protect\citeauthoryear{Schwartz et al.}{2017}]{2017MPEC....S..186S} Schwartz M. et al., 2017, 
             MPEC Circ., MPEC 2017-S186
     \bibitem[\protect\citeauthoryear{Sridhar \& Tremaine}{1992}]{1992Icar...95...86S} Sridhar S., Tremaine S., 1992, 
             Icarus, 95, 86
     \bibitem[\protect\citeauthoryear{Valsecchi, Jopek \& Froeschle}{1999}]{1999MNRAS.304..743V} Valsecchi G.~B., Jopek T.~J., Froeschle C., 1999,
             MNRAS, 304, 743
     \bibitem[\protect\citeauthoryear{Vokrouhlick{\'y} \& Nesvorn{\'y}}{2008}]{2008AJ....136..280V} Vokrouhlick{\'y} D., Nesvorn{\'y} D., 2008, 
             AJ, 136, 280
     \bibitem[\protect\citeauthoryear{Walsh, Richardson \& Michel}{2012}]{2012Icar..220..514W} Walsh K.~J., Richardson D.~C., Michel P., 2012,
             Icarus, 220, 514
  \end{thebibliography}
\end{document}